\documentclass{icrc2009}

\usepackage{graphicx}   
\usepackage[caption=false]{caption}    
\usepackage[font=footnotesize]{subfig} 
\usepackage{fixltx2e}
\usepackage{url}

\newcommand{\shorttitle}[1]%
{\markboth{Proceedings of the 31\MakeLowercase{$^{st}$} ICRC, {\L}\'{o}d\'{z} 2009}{#1} }
\newcommand{\etal}{\MakeLowercase{\textit{et al. }}} 

\begin{document}
\title{Skymap for atmospheric muons at TeV energies measured in deep-sea neutrino telescope ANTARES}

\author{\IEEEauthorblockN{Salvatore Mangano\IEEEauthorrefmark{1}, for the ANTARES collaboration}
                            \\
\IEEEauthorblockA{\IEEEauthorrefmark{1}IFIC - Instituto de F\'isica Corpuscular, Edificio Institutos de Investigati\'on, \\
                                 Apartado de Correos 22085, 46071 Valencia, Spain}}

\shorttitle{S. Mangano \etal Skymap for atmospheric muons in ANTARES}
\maketitle

\begin{abstract}
Recently different experiments mention to have observed a large
scale cosmic-ray anisotropy at TeV energies, e.g. Milagro, Tibet and Super-Kamiokande.
For these energies the cosmic-rays are expected to be nearly isotropic. Any
measurements of cosmic-rays anisotropy could bring some information about
propagation and origin of cosmic-rays.

Though the primary aim of the ANTARES neutrino telescope is the
detection of high energy cosmic neutrinos, the detector measures
mainly down-doing muons, which are decay products of cosmic-rays
collisions in the Earth's atmosphere. This proceeding describes
an anlaysis method for the first year measurement of down-going atmospheric muons at
TeV energies in the ANTARES experiment, when five out of the final
number of twelve lines were taking data.
\end{abstract}

 
\section{Introduction}

Spatial distributions of the muon flux have been measured by the Milagro
observatory \cite{Milagro}, Super-Kamiokande \cite{SuperK1, SuperK2}
as well as by the Tibet Air Shower Array \cite{Tibet}.  
These experiments report large-scale
anisotropy at primary cosmic-ray energies in the TeV range. 
The measured deviation from an isotropic distribution is of the order of $0.1\%$
and the excess region and deficit region have the size of several tens degrees. 

The ANTARES neutrino telescope \cite{Antares}, located in the Northern hemisphere, can detect
down-going muons from the North as the
above mentioned detectors. 
IceCube has also mentioned that an
investigation on a high statistic down-going muons data is ongoing.

ANTARES is located on the bottom of the
Mediterranean Sea, \mbox{40 km} off the French coast at $42^o 50'$ N,
$6^o 10'$ E.  The main objective is to detect high energy neutrinos
from galactic or extragalactic sources.  Neutrinos are detected
indirectly through the detection of Cherenkov light produced by
relativistic muons emerging from charged-current muon neutrino
interactions in the surroundings.

The detector has been successfully deployed between March 2006 and May
2008. In its full configuration the 
detector comprises 12 vertical detection lines, each of about 450 m height, 
installed at a depth of about \mbox{2500 m}. The lines
are set from each other at a distance of 60 m to 70 m. They are
anchored at the sea floor and held taut by buoys. The instrumented
part of the line starts at \mbox{100 m} above the sea floor.
Photomultipliers are grouped in triplets (up to 25 floors on each line) for a
better rejection of the optical background. They are oriented with
their axis pointing downward at an angle of $45^{\circ}$ with respect
to the vertical in order to maximize the effective area for upward-going
tracks.  ANTARES is operated in the so called all-data-to-shore mode,
which means that all photomultiplier digitized information is sent to
shore and treated in a computer farm at the shore station.  These data
are mainly due to background light caused by bioluminesence and
$^{40}K$ decay. The background light varies in time and can cause
counting rates of the photomultiplier tubes that varies between \mbox{50 kHz} 
and 500 kHz. The data flow rate at the shore station are reduced
by fast processing of the events looking for interesting physics,
which is a challenge because of the high background rates. The main
idea of the on-shore handling of these data is to take into account the
causal connection between photomultiplier tubes signals which are
compatible with the light produced by a relativistic muon.  The
reconstruction of muon tracks is based on the measurements of the
arrival times of Cherenkov photons at photomultipliers and their
positions.

Although ANTARES is optimized for upward-going particle
detection, the most abundant signal is due to the atmospheric
down-going muons. They are produced in air showers induced by
interactions of primary cosmic-rays in the Earth's atmosphere. The
muons are the most penetrating particles in such air showers. Muons
with energies above around 500 GeV can reach the detector, producing
enough amount of Cherenkov light to reconstruct the direction of the
muon.  At larger zenith angle the minimum muon energy increases. 
The muons represent a high statistic data
set that can be used for calibration purposes as well as to check the
simulation of the detector response.
The atmospheric muons will also provide information about primary cosmic rays
at energies above few TeV. For these energies the cosmic-rays
are mostly of Galactic origin and are expected to be nearly isotropic due to interactions 
with the Galactic magnetic field.   

The cosmic-ray muon spatial distribution at these energies may be not isotropic for different reasons, like the
instabilities related to temperature and pressure variations in the
Earth's atmosphere. The temperature in the upper atmosphere is related
to the density of the atmosphere and thus to the interaction of the
particles in the air showers which will affect the muon flux.
This time variations of the cosmic-ray muon flux could mimic a flux  anisotropy.
This effect is canceled out when the time scale of data taking is larger
than the time of thermodynamic changes in the Earth's atmosphere.  At
cosmic-ray energies lower of around a TeV the movements of solar
plasma through the heliosphere may change the Earth's magnetic field,
causing a modulation of the cosmic-ray anisotropy. Furthermore the
Compton-Getting effect predicts a dipole effect due to the moving of
the Earth with respect to an isotropic cosmic-ray rest system. If the
Earth is moving in the rest system, the cosmic ray flux from the
forward direction becomes larger.

\section{Skymap for down going muons}
The data used in this analysis cover the period from 
February 2007 until the middle of December 2007, during which only five
out of twelve detector lines were installed.  More than $10^7$ events
were collected, most of which are atmospheric muons. The tracks are
reconstructed with a linear fit algorithm which uses hits selected
using the time information. The algorithm used for this analysis has
the advantage to be fast, robust and has a high purity for atmospheric
muons.  The disadvantage is that this algorithms has not the ultimate
angular resolution which can be achieved by ANTARES with
a more sophisticated reconstruction algorithm.
 
Only down-going tracks detected by at least two lines and six floors
are considered for the analysis.  The angular resolution for this
selection cuts is around seven degrees.

 \begin{figure}[!t]
  \centering
  \includegraphics[width=2.5in]{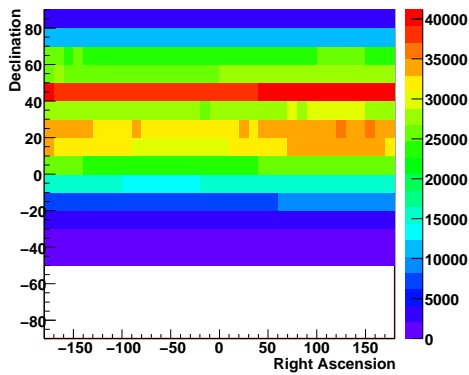}
  \caption{Muon events in equatorial coordinates for the period with
  five operating detection lines. The sky is divided into $10^{\circ}
  \times 10^{\circ}$ cells. Declinations below $-47^{\circ}$ are
  always below the horizon and are thus invisible to the detector.}
  \label{simp1_fig}
 \end{figure}

 \begin{figure}[!t]
  \centering
 \includegraphics[width=2.5in]{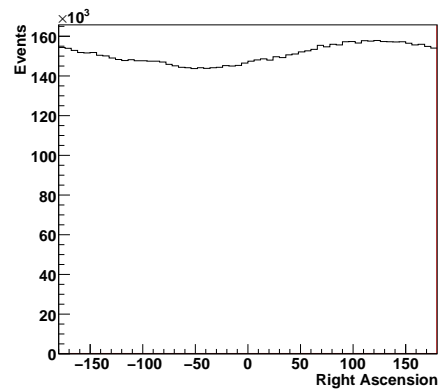}  
 \caption{Muon events as function of the right ascension without any corrections.}
  \label{simp2_fig}
 \end{figure}

The track directions are given through the zenith and azimuth angle of
the ANTARES local coordinates. Considering the time of the
reconstructed tracks, the local coordinates are translated into
equatorial coordinates: right ascension and declination. Figure
\ref{simp1_fig} shows the results of the corresponding data sample in
equatorial coordinates. The variation in declination is affected by
the visibility and by the propagation length of the muons in the
atmosphere.  The number of events as a function of the
right ascension should be uniform, because the
distribution of cosmic rays with a given zenith direction in the ANTARES local
coordinates travel a line in the declination coordinate.  A given
direction in the ANTARES local coordinates 
returns to the same right ascension after one sidereal day
because of the Earth's rotation.  The right ascension distribution is
shown in figure \ref{simp2_fig}, which is not uniform.  

The exposure for different directions is not
uniform because the detector was not constantly taking data. Several
days with high background rates have been removed for the
analysis while during normal operation the detector is   
halted regularly for calibration purposes. 
This makes the data-taking not completely uniform along the day. The
right ascension coordinates can be corrected for the introduced
fluctuations in the exposure time.  Taking the exposure time for the
analyzed runs as well as the measured direction of the track in local
coordinates, the modulation in right ascension can be calculated. The
calculated amplitude of this modulation matches well the data shown in
figure \ref{simp2_fig}. Taking these values the data right ascension
can be normalized by exposure. 

After this correction, fluctuations,
above or below the statistical expectation can be calculated. For a
given declination band with $m$ bins in a right ascension, the
probability to observe the number of events in each particular bin is
calculated.  The probability for the average expectation to fluctuate
to the observed number of events or more in this bin is calculated
with the equation
$$P(n \geq n_{obs} | \nu_b)=
1-\sum_{n=0}^{n_{obs}-1}\frac{\nu_b^n}{n!} e^{-\nu_b},$$ where $\nu_b$
is the number of expected events which is estimated from the average
background in the declination band containing that bin and $n_{obs}$ is
the observed number of events in that bin. Finally, the significance
of the cosmic ray signal of one bin is defined as
$$S=-log_{10}(P).$$

The statistics will be enlarged 
by using data of the following years. 
Assuming that the detector effects are under control, 
then the expected sensitivity to measure cosmic-ray anisotropy is 
only constrained by the number of detected muons.

\section{Conclusion}
Large scale cosmic ray anisotropies at TeV energies have been observed
by the Milagro observatory, the Tibet Air Shower Array and
Super-Kamiokande. ANTARES has a high
statistics of TeV down-going muons available, which allows to search
for possible anisotropies in the primary flux. A first attempt to
reproduce a TeV energy muon skymap is ongoing, with the one year
ANTARES data, where only five lines were deployed.

\end{document}